# Comparative Assessment of Soil-Structure Interaction Regulations of ASCE 7-16 and ASCE 7-10

Farid Khosravikia, S.M.ASCE[1]; Mojtaba Mahsuli, A.M.ASCE[2]; and Mohammad Ali Ghannad[3]

[1]PhD candidate, Dept. of Civil, Architectural, and Environmental Engineering, The University of Texas at Austin, Austin, TX, USA; e-mail: farid.khosravikia@utexas.edu
[2]Assistant professor, Dept. of Civil Engineering, Sharif University of Technology, Tehran, Iran; e-mail: mahsuli@sharif.edu
[3]Professor, Dept. of Civil Engineering, Sharif University of Technology, Tehran, Iran; e-mail: ghannad@sharif.edu

## Abstract

This paper evaluates the consequences of practicing soil structure interaction (SSI) regulations of ASCE 7-16 on seismic performance of building structures. The motivation for this research stems from the significant changes in the new SSI provisions of ASCE 7-16 compared to the previous 2010 edition. Generally, ASCE 7 considers SSI as a beneficial effect, and allows designer to reduce the design base shear. However, literature shows that this idea cannot properly capture the SSI effects on nonlinear systems. ASCE 7-16 is the first edition of ASCE 7 that considers the SSI effect on yielding systems. This study investigates the consequences of practicing the new provisions on a wide range of buildings with different dynamic characteristics on different soil types. Ductility demand of the structure forms the performance metric of this study, and the probability that practicing SSI provisions, in lieu of fixed-base provisions, increases the ductility demand of the structure is computed. The analyses are conducted within a probabilistic framework which considers the uncertainties in the ground motion and in the properties of the soil-structure system. It is concluded that, for structures with surface foundation on moderate to soft soils, SSI regulations of both ASCE 7-10 and ASCE 7-16 are fairly likely to result in a similar and larger structural responses than those obtained by practicing the fixed-base design regulations. However, for squat and ordinary stiff structures on soft soil or structures with embedded foundation on moderate to soft soils, the SSI provisions of ASCE 7-16 result in performance levels that are closer to those obtained by practicing the fixed-base regulations. Finally, for structures on very soft soils, the new SSI provisions of ASCE 7-16 are likely to rather conservative designs.

# Introduction

This paper aims at assessing the consequences of practicing soil structure interaction (SSI) regulations of the 2016 version of ASCE 7 (American Society of Civil Engineers, 2016), hereafter referred to as ASCE 7-16, on seismic performance of building structures. Generally, the basis of SSI provisions in ASCE is to replace the soil-structure system with an equivalent fixed base model with a longer period and usually, a larger damping ratio. Thus, designers are allowed to reduce the design base shear in the equivalent lateral force procedure. Although this idea works for linear soil-structure systems, the literature shows that it cannot properly capture the SSI effects on nonlinear systems. In fact, there are two main challenges in evaluating the effect of SSI on yielding systems. First, it has been shown that SSI is less influential on yielding systems, and the effect of SSI diminishes as the structure experiences more nonlinearity (Veletsos and Verbic 1974). Second, it has been shown that SSI may lead in larger structural responses, which may adversely affect the performance of the buildings (Barcena and Esteva 2007; Bielak 1978; Ghannad and Ahmadnia 2006; Tang and Zhang 2011). The effect of SSI on the structural performance of the yielding structures was more recently addressed by Khoshnoudian et al. (2013), Demirol and Ayoub (2017), and Hassani et al. (2018). Moreover, taking one step further, Khosravikia (2016) and Khosravikia et al. (2018) showed that SSI may increase the seismic risk of buildings. Therefore, assuming a purely beneficial effect for SSI on yielding structures, which is the basis of the seismic design codes, needs further investigations.

ASCE 7-16 is the first edition of ASCE 7 that takes into account the SSI effect on yielding systems by introducing a cap for the base shear reduction as a function of the expected level of nonlinearity in the structure. These provisions are built upon the SSI provisions of the 2015 NEHRP (FEMA 2015), which have been scrutinized by the authors in another work (Khosravikia et al. 2017). ASCE 7-16 recommends smaller reductions of the design base shear for systems with larger nonlinear deformation capacity. Even though this is a welcome change, the consequences of practicing the new provisions should be studied quantitatively, which forms the main motivation of this research. The results obtained by practicing SSI regulations of ASCE 7-16 are compared with those from 2010 edition of ASCE 7 (2010), hereafter referred to as ASCE 7-10, to assess if the new ASCE 7-16 SSI provisions can be considered an improvement upon the previous SSI provisions of ASCE 7-10.

The ductility demand of the structure forms the metric of the structural performance in this study. The soil-structure system is here modeled by the sub-structure method in which the structure is modeled by a single-degree-of-freedom system with an idealized bilinear behavior. The soil is considered as a homogenous half-space and replace by a discrete model based on the concept of cone models (Meek and Wolf 1994). The proposed framework takes into account the prevailing uncertainties in the ground motion and in the properties of the soil-structure system. The uncertainties in the soil and the structure are described by random variables that are input to soil-structure model, and the uncertainty in the ground motion is considered by using a suite of over 6,000 ground motions recorded on soil. The analyses are conducted on a large number of soil-structure systems with different numbers of stories, structural systems, aspect ratios, and foundation embedment ratios located on different site classes. For each system, a Monte Carlo sampling analysis produces the probability that practicing SSI provisions, in lieu of fixed-base provisions, increases the ductility demand of the structure. The probability distributions of the ductility demand of the structure that is designed once in accordance with the SSI provisions of ASCE 7-16 and once using the conventional fixed-base provisions are computed and compared.

## Soil-Structure Model

The soil-structure systems considered in this study are modeled using the sub-structure method. This study takes into account two main effects of SSI, namely inertial and kinematic interactions. The inertial interaction refers to the interaction in which the flexibility of the soil beneath the structure affects the structural responses of the buildings by affecting its dynamic characteristics. The flexibility of the soil causes a longer natural period for the building, and the contribution of the material damping as well as the radiation damping of the soil generally leads to a higher damping ratio for the new system. To properly take into account the inertial effect of the SSI, as shown in Figure 1, the soil-structure systems are modeled as a system with 4 degrees of freedom (DOF). The first DOF is to simulate the structure deformation, and the other three DOFs are to simulate sway and rocking DOFs of the foundation as well as frequency dependency of the soil dynamic stiffness. Each considered DOF of the soil-structure system is discussed in the following paragraphs.

As noted, the first DOF of the soil-structure system, which is represented by $u$ in Figure 1, is the structure deformation. In fact, the super-structure is simulated by a single-degree-of-freedom (SDOF) system with mass $m$, stiffness $k$, damping coefficient $c$, mass moment of inertia $I$, and height $h$. It should be noted that this DOF represents the first mode of vibration of a multi-degree-of-freedom (MDOF) system, and the nonlinearity of the buildings is taken into account by assigning an elastic, perfectly plastic behavior for the spring. The two parameters that define this behavior are the stiffness and yield strength of the structure. The former is computed from the period of the structure, and the latter is computed in accordance with the conventional fixed-base provisions of ASCE-7. That is, a base shear coefficient is computed for each system given its period, $T_n$, response modification factor, $R$, and the standard design spectrum associated with specific values of the spectral acceleration at short periods, $S_S$, and at the period of 1.0s, $S_1$. Here, values of 1.25 and 0.6 are respectively assigned for $S_S$ and $S_1$, which represents a highly seismically-active region. It is worth noting that the response modification factor determines the level of nonlinearity in the structure.

Two sway and rocking degrees of freedom are to simulate the behavior of the foundation, which is here assumed to be a rigid cylinder, embedded in the half-space soil. The main characteristics of the foundation are mass, $m_f$, mass moment of inertia, $I_f$, radius, $r$, and embedment depth, $e$. The sway and rocking stiffness of the foundation, which are respectively shown by $k_h$ and $k_r$, are as follows:

$$k_h = \frac{8\rho v_s^2 r}{2-v}\left(1+\frac{e}{r}\right), \ k_r = \frac{8\rho v_s^2 r^3}{3(1-v)}\left[1+2.3\frac{e}{r}+0.58\left(\frac{e}{r}\right)^3\right] \tag{1}$$

where $\rho$, $v$, and $v_s$ respectively refer to the density, Poisson's ratio, and shear wave velocity of the soil, and $r$ is radius of the cylindrical foundation. To account for the nonlinearity of the soil, an equivalent elastic approach is used (Kramer 1996). In this approach, the stiffness of sway and rocking springs $k_h$ and $k_r$ are computed using a degraded shear wave velocity that depends on the level of strains in soil. Here, the degradation factors suggested by ASCE-7 (2016) is used to model the nonlinearity of the soil, which are discussed in the following section. The sway and rocking damping ratios of the foundation are as follows:

$$c_h = \frac{r}{v_s}\gamma_{0h}k_h, \ c_r = \frac{r}{v_s}\gamma_{0r}k_r \tag{2}$$

where $\gamma_{0h}$ and $\gamma_{0r}$ are dimensionless parameters that are computed given the foundation embedment ratio, $e/r$ (Wolf 1994). The sway and rocking dashpots model the radiation damping of the soil beneath the structure. It should be noted that the sway spring and dashpot are connected to the foundation with eccentricities $f_k$ and $f_c$, which can be computed as:

$$f_k = 0.25e, \quad f_c = 0.32e + 0.03e\left(\frac{e}{r}\right)^2 \tag{3}$$

Finally, the last DOF, known as the monkey tail in Wolf (1994), simulates the frequency dependency of the soil dynamic stiffness. The monkey tail, which simplifies modeling of other coefficients independent of the excitation frequency, comprises a mass and a dashpot with the following coefficients:

$$c_m = \frac{r}{\upsilon_s}\gamma_{1r}k_r, \quad I_m = \left(\frac{r}{\upsilon_s}\right)^2 \mu_{1r}k_r \tag{4}$$

where $\gamma_{1r}$ and $\mu_{1r}$ are two more dimensionless coefficients in terms of $e/r$ according to Wolf (1994).

Generally, there are three sources for the damping in a soil-structure system: 1) the damping of the building, which is considered by assigning a dashpot for the building; 2) soil radiation damping which is considered by sway and rocking dashpots shown in Figure 1a; 3) soil material damping, which is introduced in the model using the correspondence principle (Wolf 1985) in which each original spring in the discrete model is accompanied by a dashpot, and each original dashpot by a mass. The coefficients of these dashpots and masses are demonstrated in Figure 1b.

The second effect of the SSI is the kinematic interaction. A rigid embedded foundation turns a single horizontal component of the free-field motion into two sway and rocking components. These two components that are the responses of the massless rigid foundation to the free-field motion form the foundation input motion. In this study, these two components are computed using the method suggested by Meek and Wolf (1994). The four-DOF discrete model in Figure 1 is then subjected to these two components. The analyses are conducted in time domain using the Newmark β method (Newmark 1959). In each time step, the average acceleration with the key parameters of β=0.25 and γ=0.5 is employed, so the analysis is unconditionally stable (Chopra 2012).

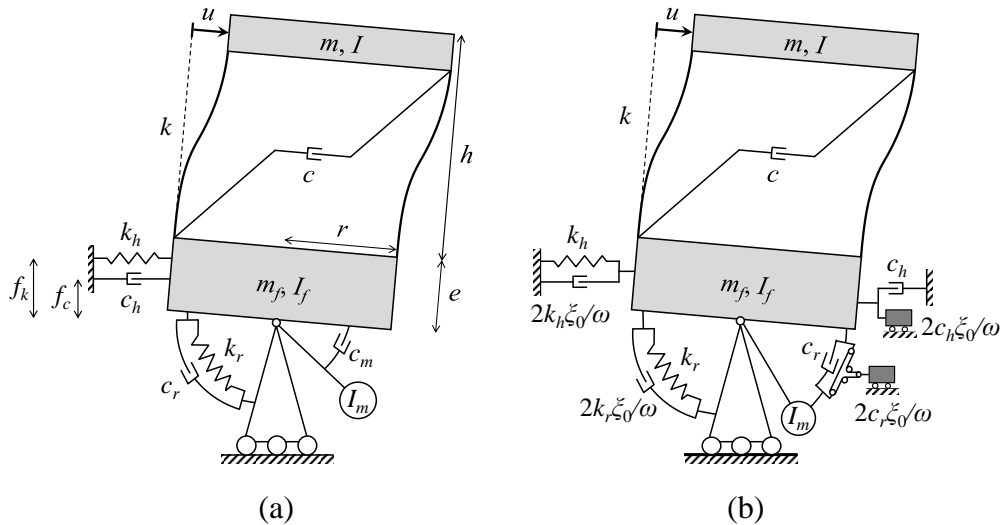

Figure 1: (a) Basic soil-structure model; (b) model with soil material damping

## Soil-Structure Systems

The systems considered in this study are the same as the ones previously employed by khosravikia et al. (2017). They considered 720 soil-structure systems with different numbers of stories, structural systems, aspect ratios, and foundation embedment ratios on various site classes. This section gives a brief introduction on the considered soil-structure systems. Generally, the dynamic behavior of the soil-structure systems are mainly controlled by three dimensionless parameters (Ghannad et al. 1998; Veletsos 1977): (1) a dimensionless parameter, $a_0$, that represents the relative stiffness of the structure to that of the soil; (2) aspect ratio of the building, $h/r$, that represents the ratio of the effective height of the SDOF structure to the radius of the foundation; and (3) the embedment ratio of the foundation, $e/r$.

The dimensionless frequency, $a_0$, is an indicator of the severity of the SSI effects and can be computed using the following equation:

$$a_0 = \frac{\omega_n h}{\upsilon_S} \tag{5}$$

where $\omega_n$=natural circular frequency of the fixed-base structure. As seen, this parameter depends on the natural period of the fixed base system and shear wave velocity of the soil. In this study, the fixed-base natural period of the structure is estimated using the relationship from ASCE-7 (2016), i.e., $T_n=C_t \cdot h_n^x$. In this equation, $C_t$ and $x$ are coefficients determined based on the structural system, and $h_n$ is the total height of the structure. In this study, in order to consider low to high levels of SSI effect, i.e., wide range of $a_0$ values, buildings with different structural systems and heights located on different soil types are considered. In particular, the present study considers two types of structure: First, *Stiff* structures with a lower-bound period in which $C_t=0.0488$ and $x=0.75$; Second, *Soft* structures with an upper-bound period in which $C_t=0.0724$ and $x=0.8$. The response modification factor which represent the level of nonlinearity in the structure is assumed to be 6 for *Stiff* structures and is assumed to be 8 for *Soft* structures. Examples of the former are steel special concentrically braced frame systems or special reinforced concrete shear wall systems, and examples for the latter include steel special moment-resisting frame systems. For each type of structure, buildings with 20 different heights associated with 20 different number of stories, $N_{story}$, are considered. In fact, for each type of structure, $N_{story}$ varies from 1 to 20, and for each $N_{story}$, it is assumed that the total height of the building is equal to $3.2N_{story}$.

Then, each building are assumed to be located on three different soil types as Site Classes C, D, and E according to ASCE-7 (2016) site classification, which represent moderate to very soft soils. For each soil type, as mentioned earlier, the degradation factors suggested by ASCE-7 are employed to degrade the shear wave velocity to account for the nonlinearity of soil. According to ASCE-7, the degradation factor for shear wave velocity depends on the effective peak acceleration, $S_{DS}/2.5$, which depends on the values of the spectral acceleration at short periods, $S_S$, and at the period of 1.0s, $S_1$. As noted earlier, values of 1.25 and 0.6 are respectively assigned for the values of $S_S$ and $S_1$, which results in the degradation factors, of 0.89, 0.77, and 0.35 for Site Classes C, D, and E, respectively.

In order to represent conventional buildings, the mean of $h/r$ takes on the values of 0.5, 1, and 2, which represent squat, ordinary, and slender structures, respectively. The uncertainty of this parameter is taken into account by the method suggested with Khosravikia (2016) and Khosravikia et al. (2017). Moreover, the constant values of 0, 0.5, 1.0 are considered for the embedment ratio of the foundation, $e/r$, to represent surface and embedded foundations. The uncertainty of this parameter is neglected in this study. It is worth noting that for squat structures no embedded foundation is considered in this study, and the embedment ratio of 1.0 is only taken into account for slender structures. As a result, six different combinations of *h/r* and *e/r* are considered in this study.

Combination of the abovementioned parameters, i.e. the two types of structural systems (*Stiff* and *Soft*), three Site Classes (C, D, and E), six combinations of aspect and embedment ratios, and 20 different numbers of stories, results in 720 different soil-structure systems. Each of these systems is modeled using the four-DOF model of Figure 1. In addition to above-mentioned key parameters, the structure-to-soil mass ratio index, $\bar{m}=m/\rho r^2 h$, the foundation-to-structure mass ratio, $m_f/m$, the material damping ratio of the soil, $\xi_g$, the material damping ratio of the structure, $\xi_s$, and Poisson's ratio of the soil, $v$, are employed to describe the other properties of the soil-structure system. To properly account for the uncertainty of these parameters, they are modeled as random variables with probability distributions that are derived by the authors in previous studies; see Khosravikia (2016) and Khosravikia et al. (2017).

The soil-structure model and analysis methods are implemented in Rt, which is a computer program for multi-model reliability and risk analysis (Mahsuli and Haukaas 2013). Rt has an object-oriented design for probabilistic analysis, including reliability, sensitivity, and optimization analysis with multiple interacting probabilistic models.

## Probabilistic Evaluation Metric

ASCE and other current seismic design codes generally consider SSI as a purely beneficial effect. Therefore, they provide engineers with two acceptable design methodologies. The first one is the conventional method in which the structure is designed as a fixed-base system, i.e., the soil under the structure is assumed to be rigid. There is a wealth of knowledge on the behavior and the performance of fixed-base systems, which forms the basis of setting the target level of safety in design codes. The second one, which is optional, allows the designer to reduce the design base shear for soil-structure interacting systems, hereafter referred to as SSI systems. It is expected that the two methodologies yield acceptable performance. In this study, the probability that the ductility demand of the structure designed per SSI provision of ASCE, $\mu_{SSI}$, exceeds that of the structure designed per fixed-base provisions, $\mu_{fix}$, i.e., $P(\mu_{SSI}>\mu_{fix})$, is used as the metric for the performance evaluation. This metric was first introduced by the authors in Khosravikia et al. (2017). In that study, this metric was used to evaluate the consequences of practicing SSI provisions of ASCE 7-10 and those of NEHRP, which forms the basis of the new SSI provisions of ASCE.

The values of $P(\mu_{SSI}>\mu_{fix})$ for different systems are computed using Monte Carlo sampling reliability analysis. For each system, 20,000 samples are generated to achieve an accurate prediction of $P(\mu_{SSI}>\mu_{fix})$. In each sample, each system is designed according to both methodologies. Each system is then subjected to a randomly selected ground motion, and nonlinear dynamic time history analyses are conducted to evaluate the consequences of the design methodologies. Considering 20,000 samples in the Monte Carlo sampling analysis of each pair of soil-structure and fixed-base systems, 31.2 million nonlinear dynamic time history analyses are conducted to estimate the target probability for all systems. This massive study is conducted by parallel processing on the cluster of the High-Performance Computing Center at Sharif University of Technology. To comprehensively model the ground motion uncertainty, a suite of 6848 accelerograms recorded on soil is employed. The records are obtained from the strong motion database of the Pacific Earthquake Engineering Research Center (Chiou et al. 2008). In order to investigate the performance of the systems at the design level, the randomly selected record is scaled to the design spectral acceleration at the period of the structure, $S_a$, from the code. For more detail about the ground motions and the scaling method, see Khosravikia et al. (2018). Different ranges for the values of $P(\mu_{SSI}>\mu_{fix})$ represent different interpretation on the consequences of practicing SSI provisions, as follows:

1) $P(\mu_{SSI}>\mu_{fix}) < 0.4$ represent the cases in which the SSI design procedure results in conservative and uneconomic design compared to the fixed design procedure. In this case, more reduction in the design base shear proposed by the code can be considered.
2) $0.4 \leq P(\mu_{SSI}>\mu_{fix}) \leq 0.6$ represent the cases in which the SSI design procedure results in nearly the same median performance as that of the fixed-base design procedure. In this case, the amount of the reduction in the design base shear proposed by the code is considered to be optimal.
3) $P(\mu_{SSI}>\mu_{fix}) > 0.6$ represent the cases in which the SSI design procedure results in unsafe level of performance compared to the fix based design procedure. In such cases, the reduction in the design base shear proposed by the code is excessive, and the beneficial effect of SSI cannot counteract it.

Figure 2 shows $P(\mu_{SSI}>\mu_{fix})$ against the number of stories for *Soft* structures. This figure compares the abovementioned probability for 360 soil-structure systems that are once designed in accordance with ASCE 7-10 and once with ASCE 7-16. The probability is demonstrated for various site classes, aspect ratios, and embedment ratios. For detail discussion about the figure, one may visit Khosravikia at al. (2017). It should be noted that the gray background refers to unconventional systems given the aspect ratio. As seen in the figure, practicing the SSI provisions of the two design codes results in different level of performance for different systems, which is correlated with the amount of design base shear reduction proposed by the corresponding SSI provisions.

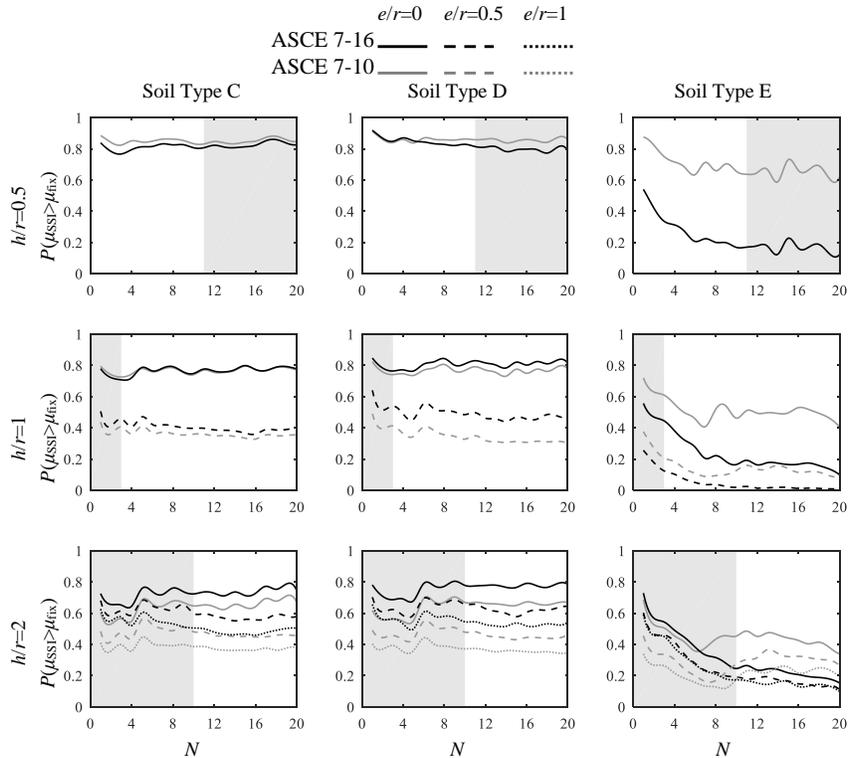

Figure 2: $P(\mu_{SSI}>\mu_{fix})$ from ASCE 7-16 and ASCE 7-10 for *Soft* structures

## Comparison of SSI Provisions of ASCE 7-16 and ASCE 7-10

In this section, the consequences of practicing the two generations of SSI provisions are compared using the metric defined in the previous section. To do so, Figure 3a and Figure 4a present the comparative assessment between SSI provisions of ASCE 7-10 and ASCE 7-16 for *Stiff* and *Soft* structures, respectively. Each cell in these figures characterizes a particular soil-structure system. The properties of each structure and site class are shown on left and the number of stories is shown on the bottom. It should be noted that the crossed-out cells refer to the unconventional systems considering the number of stories and the given aspect ratio. The color in each cell visually illustrates one of the following cases:

- Case 1, represented by blue color in the figures, refers to the cases that SSI provisions of ASCE 7-16 can be considered as an improvement upon those of ASCE 7-10. An improvement, in this context, means that $P(\mu_{SSI} > \mu_{fix})$ from ASCE 7-16 is at least 0.1 closer than $P(\mu_{SSI} > \mu_{fix})$ from ASCE 7-10 to the reference probability of 0.5, which can be computed as follows:

$$\left| P(\mu_{SSI} > \mu_{fix}) - 0.5 \right|_{ASCE\ 7\text{-}16} - \left| P(\mu_{SSI} > \mu_{fix}) - 0.5 \right|_{ASCE\ 7\text{-}10} \leq -0.1 \quad (6)$$

- Case 2, represented by yellow color in the figures, refers to the cases that SSI provisions of ASCE 7-16 make insignificant changes in the structural performance compared to those of ASCE 7-10. In fact, the new provisions change $P(\mu_{SSI} > \mu_{fix})$ less than 0.1, which can be computed as follows:

$$-0.1 < \left| P(\mu_{SSI} > \mu_{fix}) - 0.5 \right|_{ASCE\ 7\text{-}16} - \left| P(\mu_{SSI} > \mu_{fix}) - 0.5 \right|_{ASCE\ 7\text{-}10} < 0.1 \quad (7)$$

- Case 3, represented by brown color in the figures, refers to the cases that SSI provisions of ASCE 7-16 may not be considered as an improvement upon those of ASCE 7-10. In this case, $P(\mu_{SSI} > \mu_{fix})$ from ASCE 7-16 is at least 0.1 farther than $P(\mu_{SSI} > \mu_{fix})$ from ASCE 7-10 from the reference probability of 0.5, which can be computed as follows:

$$\left| P(\mu_{SSI} > \mu_{fix}) - 0.5 \right|_{ASCE\ 7\text{-}16} - \left| P(\mu_{SSI} > \mu_{fix}) - 0.5 \right|_{ASCE\ 7\text{-}10} \geq 0.1 \quad (8)$$

In this comparison, when it is notably likely that the SSI provisions lead to unsafe designs, i.e., when $P(\mu_{SSI} > \mu_{fix}) > 0.6$, the difference of such probabilities with the reference value of 0.5 is doubled. This is owing to the fact that conservative design is more preferable than unsafe design for engineers.

Moreover, Figure 3b and Figure 4b show the details of each case for ASCE 7-16. Again, the plots at top, middle, and bottom show the results for the structures on Site Class C, D, and E, respectively. For each case, the red color represents the number of the samples, *N*, for which practicing SSI provision of ASCE 7-16, regardless of the results for ASCE 7-10, results in unsafe design compared to the fixed-base provisions. For these systems, $P(\mu_{SSI} > \mu_{fix})$ from ASCE 7-16 is higher than 0.6. The green color represents the number of samples for which the new SSI provisions yield a conservative design for the soil-structure system compared to the fixed-base design methodology. For these systems, $P(\mu_{SSI} > \mu_{fix})$ from ASCE 7-16 is less than 0.4. Finally, the white part shows the number of the samples that SSI provisions of ASCE 7-16 lead to optimal design, which means that practicing those provisions results in almost the same level of performance as the fixed-base design methodology. For these systems, $P(\mu_{SSI} > \mu_{fix})$ from ASCE 7-16 is between 0.4 and 0.6.

The white color for structures on Site Class C in Figure 3a and Figure 4a show that the new

SSI provisions generally bring about insignificant changes in the structural performance compared to those of ASCE 7-10. The abundance of the red color for such systems, as shown by Case 3 in Figure 3b and Figure 4b, uncovers that it is mostly likely that both provisions result in unsafe designs for such systems. Therefore, in order to achieve an optimal design for such structures, the base shear reduction should be limited even further. However, for buildings with deeply-embedded foundations on this site class, the new SSI provisions of ASCE 7-16 show an improvement. The white color for such systems, as shown by Case 1 in Figure 3b and Figure 4b, reveals that this improvement results in optimal designs for those buildings with deeply-embedded foundations.

The brown and white colors for structures on Site Class E in Figure 3a and Figure 4a show that the new SSI provisions ASCE 7-16 cannot be considered as an improvement upon the previous SSI provisions for these structures. In fact, the white color in Figure 3a denotes that the new SSI provisions lead to the similar structural performance as those of ASCE 7-10 for *stiff* structures on Site Class E, and the brown color in Figure 4a shows that the previous SSI provisions result in better estimation of the performance of *Soft* structure compared to the new provisions. In addition, the green hue in the right-side plots, i.e. Figure 3b and Figure 4b, shows that ASCE 7-16 provisions generally result in overly un-conservative designs for such structures. The reason is that SSI is very beneficial for such structures, and the proposed reductions from ASCE are not enough to fully take advantage of this beneficial effect (Khosravikia et al. 2017). Hence, further reductions in the design base shear proposed by ASCE 7-16 could be considered.

Finally, the consequences of the SSI provisions on buildings located on Site Class D are discussed. First, the consequences of the SSI provisions on structures with surface foundation on this site class are investigated. As seen in Figure 4a, the new SSI provisions cannot be considered as an improvement upon the previous SSI provisions for *Soft* structures with surface foundation. The red hue, as shown by Case 2 and Case 3 in Figure 4b, reveals that the new SSI provisions lead to unsafe designs for such systems. However, as shown in Figure 3a, the new provisions have a better estimation for *Stiff* structures on this site class compared to the previous SSI provisions. In fact, the previous SSI provisions result in unsafe designs for squat and ordinary-size *Stiff* structures; however, the new SSI provisions allows much less reduction in the design base shear compared to those of ASCE 7-10, which results in better design solutions (Khosravikia et al. 2017). In particular, the white hue, as shown by Case 1 in Figure 3b, shows that the new provisions lead to optimal designs for squat and ordinary-size *Stiff* structures. Next, the consequences of the SSI provisions on structures with embedded foundation on this site class are investigated. For buildings with embedded foundations, the SSI provisions of ASCE 7-16 usually have a better estimation of the performance of the structure. The white hue for such systems, as shown by Case 1 in Figure 3b and Figure 4b, reveals that this improvement results in optimal designs for buildings with embedded foundation on this site class.

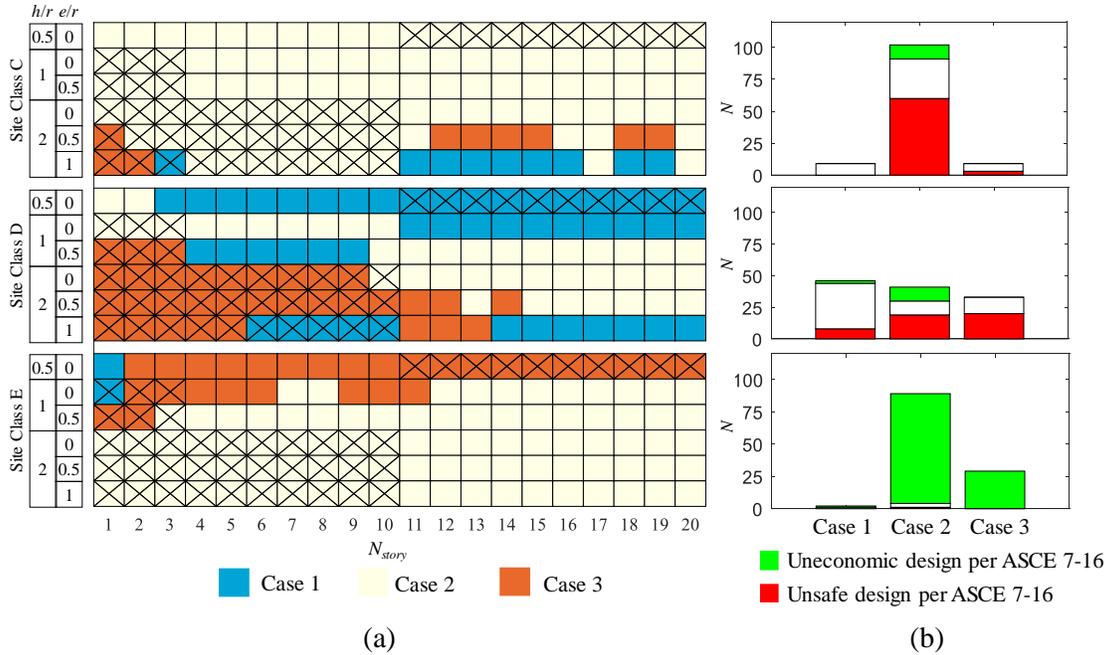

Figure 3: Comparative assessment of SSI provisions of ASCE 7-16 with those of ASCE 7-10 for *Stiff* Structures

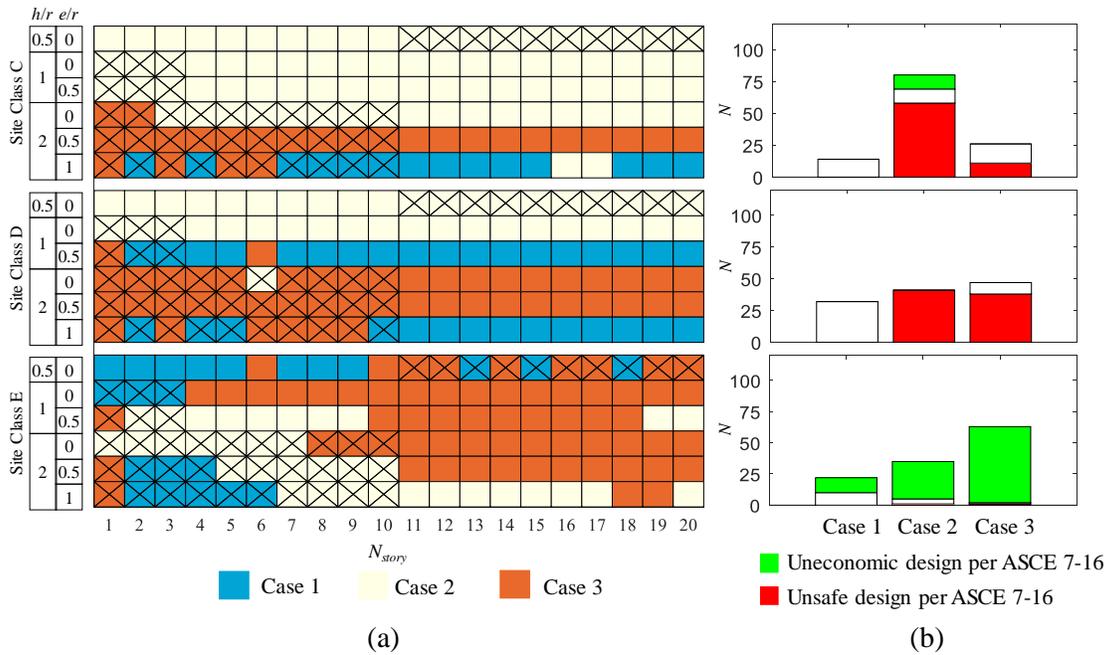

Figure 4: Comparative assessment of SSI provisions of ASCE 7-16 with those of ASCE 7-10 for *Soft* structures

## Conclusions

This paper investigates the consequences of practicing soil structure interaction (SSI) regulations of ASCE 7-16 on seismic performance of building structures. In particular, it is uncovered if the amount of design base shear reduction suggested by ASCE 7-16 is optimal for a wide range of buildings with different numbers of stories, structural systems, aspect ratios, and foundation embedment ratios located on different site classes. For this purpose, a probabilistic metric is used, which is defined as the probability that the ductility demand of the structure designed per SSI provisions of ASCE exceeds that of the structure designed per conventional fixed-base provisions. In this context, optimal design base shear reduction is defined as the reduction that results in nearly the same median structural performance of the building as that of the conventional fixed-base design procedure, which is the target level of safety in the design codes. Then, it is investigated if these provisions can be considered as an improvement upon the previous 2010 edition, i.e. the new provisions suggest more optimal values for the design base shear of the buildings. The major findings of this study are as follows:

- For structures on Site Class C, it is fairly likely that practicing SSI provisions of ASCE 7-16, similar to those of ASCE 7-10, result in larger structural responses compared to those obtained by practicing the corresponding fixed-base structures. Hence, a smaller amount of reduction in the design base shear is suggested by the authors for these systems. However, for buildings with deeply-embedded foundations on this site class, the new SSI provisions of ASCE 7-16 have a better estimation of the structural performance. In fact, for such systems, SSI provisions of ASCE 7-16 result in performance levels that are closer to those obtained by the fixed-base regulations.
- For most structures with surface foundation on Site Class D, SSI provisions of ASCE 7-16, by and large, cannot be considered as an improvement upon the previous SSI provisions. However, for squat or ordinary-size *Stiff* structures, SSI provisions of ASCE 7-16 have a better estimation of the performance of the structure compared to those of ASCE 7-10. For such systems, ASCE 7-16 proposes more optimal values for the design base shear reduction compared to the previous SSI provisions.
- for structures with embedded foundation on Site Class D, SSI provisions of ASCE 7-16 can be considered as an improvement upon those of ASCE 7-10 and lead to optimal designs. In fact, the suggested amount of reduction in the design base suggested by ASCE 7-16 is optimal for such systems.
- For structures on Site Class E, ASCE 7-16 provisions are generally likely to result in overly conservative designs. For such structures, more reduction in the design base shear could be considered.

It should be noted that the reliability and accuracy of the results, in this study, are limited to the structures on moderate to very soft soils in highly seismically-active regions, as well as to the other modeling assumptions, i.e., structures with an elastic, perfectly plastic behavior and with rigid, mat foundations on or embedded in the soil half-space.

# References

American Society of Civil Engineers,. (2010). *Minimum design loads for buildings and other structures*. ASCE/SEI 7, Reston, VA.

American Society of Civil Engineers,. (2016). *Minimum design loads and associated criteria for buildings and other structures*. ASCE/SEI 7, Reston, VA.

Barcena, A., and Esteva, L. (2007). "Influence of dynamic soil-structure interaction on the nonlinear response and seismic reliability of multistorey systems." *Earthquake engineering & structural dynamics*, Wiley Online Library, 36(3), 327–346.

Bielak, J. (1978). "Dynamic response of non-linear building-foundation systems." *Earthquake

*Engineering & Structural Dynamics*, John Wiley & Sons, Ltd, 6(1), 17–30.

Chiou, B., Darragh, R., Gregor, N., and Silva, W. (2008). "NGA project strong-motion database." *Earthquake Spectra*, 24(1), 23–44.

Chopra, A. (2012). *Dynamics of Structures: Theory and Applications to Earthquake Engineering*. Prentice Hall, New Jersey.

Demirol, E., and Ayoub, A. S. (2017). "Inelastic displacement ratios of SSI systems." *Soil Dynamics and Earthquake Engineering*, Elsevier, 96, 104–114.

FEMA. (2015). *NEHRP recommended seismic provisions for new buildings and other structures*. FEMA P-1050, Washington, DC.

Ghannad, M. A., and Ahmadnia, A. (2006). "The effect of soil-structure interaction on inelastic structural demands." *European Earthquake Engineering*, 20(1), 23–35.

Ghannad, M. A., Fukuwa, N., and Nishizaka, R. (1998). "A study on the frequency and damping of soil-structure systems using a simplified model." *Journal of Structural Engineering, Architectural Institute of Japan (AIJ)*, 44(B), 85–93.

Hassani, N., Bararnia, M., and Amiri, G. G. (2018). "Effect of soil-structure interaction on inelastic displacement ratios of degrading structures." *Soil Dynamics and Earthquake Engineering*, Elsevier, 104, 75–87.

Khoshnoudian, F., Ahmadi, E., and Nik, F. A. (2013). "Inelastic displacement ratios for soil-structure systems." *Engineering Structures*, Elsevier, 57, 453–464.

Khosravikia, F. (2016). "Seismic risk analysis considering soil-structure interaction." Master's Thesis, Department of Civil Engineering, Sharif University of Technology, Tehran, Iran.

Khosravikia, F., Mahsuli, M., and Ghannad, M. A. (2017). "Probabilistic evaluation of 2015 NEHRP soil-structure interaction provisions." *Journal of Engineering Mechanics*, 143(9), 04017065.

Khosravikia, F., Mahsuli, M., and Ghannad, M. A. (2018). "The effect of soil–structure interaction on the seismic risk to buildings." *Bulletin of Earthquake Engineering*, DOI: 10.1007/s10518-018-0314-z.

Kramer, S. L. (1996). *Geotechnical earthquake engineering*. Prentice Hall, New Jersey.

Mahsuli, M., and Haukaas, T. (2013). "Seismic risk analysis with reliability methods, part II: Analysis." *Structural Safety*, 42, 63–74.

Meek, J. W., and Wolf, J. P. (1994). "Cone models for embedded foundation." *Journal of Geotechnical Engineering*, 10.1061/(ASCE)0733-9410(1994)120:1(60), 60–80.

Newmark, N. M. (1959). "A method of computation for structural dynamics." *Journal of the Engineering Mechanics Division*, 85(3), 67–94.

Tang, Y., and Zhang, J. (2011). "Probabilistic seismic demand analysis of a slender RC shear wall considering soil-structure interaction effects." *Engineering Structures*, 33(1), 218–229.

Veletsos, A. S. (1977). "Dynamics of structure-foundation systems." *Structural and geotechnical mechanics*, W. J. Hall, ed., Prentice-Hall, New Jersey, 333–361.

Veletsos, A. S., and Verbic, B. (1974). "Dynamics of elastic and yielding structure-foundation systems." *5th World Conference on Earthquake Engineering*, Editrice Libraria, Italy.

Wolf, J. P. (1985). *Dynamic soil-structure interaction*. Prentice Hall, Upper Saddle River, New Jersey.

Wolf, J. P. (1994). *Foundation vibration analysis using simple physical models*. Prentice Hall, New Jersey.